\documentclass[aps,prl,twocolumn,showpacs,groupedaddress]{revtex4}
\usepackage{graphicx}  % needed for figures
\usepackage{dcolumn}   % needed for some tables
\usepackage{bm}        % for math
\usepackage{amssymb}   % for math
\usepackage{amsmath}
\usepackage{color}
\usepackage{bbold}

\begin{document}

% The following information is for internal review, please remove them for submission
%%\widetext
%%\leftline{Version xx as of \today}
%%\leftline{Primary authors: Joe E. Physics}
%%\leftline{To be submitted to (PRL, PRD-RC, PRD, PLB; choose one.)}
%%\leftline{Comment to {\tt d0-run2eb-nnn@fnal.gov} by xxx, yyy}
%%\centerline{\em D\O\ INTERNAL DOCUMENT -- NOT FOR PUBLIC DISTRIBUTION}

% the following line is for submission, including submission to the arXiv!!
%\hspace{5.2in} \mbox{Fermilab-Pub-04/xxx-E}
%\author{D. Avisar and D.J. Tannor \\ \textit{Chemical Physics
%Department, Weizmann Institute of Science, Rehovot, Israel}}

\title{Thermal-Light-Induced Coherent Dynamics in Atoms and Molecules --
\\an Exact Quantum Mechanical Treatment}
%\input author_list.tex       % D0 authors (remove the first 3 lines
                             % of this file prior to submission, they
                             % contain a time stamp for the authorlist)
                             % (includes institutions and visitors)

%
\author{David Avisar$^{\dagger}$} %\affiliation{Chemical Physics Department,

\author{Asaf Eilam} %
\author{A.D. Wilson-Gordon}

\affiliation{Department of Chemistry, Bar-Ilan University,
 Ramat Gan 5290002, Israel}
\vskip 0.25cm

\date{\today}

\begin{abstract}

The question of whether sunlight induces coherent dynamics in biological
systems is under debate. Here we show, on the basis of an exact fully quantum mechanical treatment, that thermal light
induces excited-state coherences in matter similar to those induced by a coherent state.
We demonstrate the phenomenon on a $V$-type model system and a two-state Born-Oppenheimer
molecular system. Remarkably, wavepacket-like dynamics is induced in the excited molecular potential-energy surface.

\end{abstract}

\maketitle

%\section{\label{sec:level1}First-level heading}
% sections are not used for PRL papers

Coherence is a fundamental and central theme in quantum mechanics and is
manifested in light-matter interaction through a variety
of unique phenomena \cite{scully_book}, including quantum control \cite{shapiro_book},
storage and retrieval of information \cite{lukin}, and quantum state reconstruction \cite{phil, avisar}.
In recent years, there has been a lively debate on whether
sunlight-induced chemical reactions in biological systems evolve
coherently and consequently exhibit extremely efficient energy transfer.
Experimental evidence for vibrational and electronic coherence in such systems
has been reported in several studies \cite{shank_1, martin, shank_2, shank_3, fleming, fleming_2, engel, scholes}.
However, it has been argued that
since such studies employ coherent laser pulses rather than
sunlight (so-called thermal or chaotic light) to initiate material dynamics,
the results obtained cannot establish conclusively
that sunlight-induced biological processes evolve coherently \cite{shapiro}.

Theoretical studies of the excitation of atomic and molecular
systems by thermal light have reached contradictory conclusions as to whether
the induced dynamics is coherent. Most of these studies are
approximate in some sense, as they involve either semiclassical or perturbative
treatments, and some of them rely on the presence of an intrinsic
coupling between the material excited states,
such as vacuum- or decay-induced coherence
\cite{brumer_3, scully_3, plenio_2, scully, scully_2, brumer_2, brumer, gordon}.
Others, treat the material
system only following an assumed coherent excitation \cite{plenio}.
None of these studies employ an exact and fully quantum mechanical treatment of
the interaction of thermal light with matter without such limiting constraints.
Interestingly, the entanglement (correlations) between separated material
sites has also been studied in this context \cite{birgitta, mukamel}.

Since the interaction of thermal light with matter
is a fundamental issue of enormous potential
implications, the need for a definite conclusion on whether
thermal light can induce coherent material dynamics is of
primary importance and significance.
In order to reach a definite conclusion, we treat the
problem starting from basic considerations which,
to the best of our knowledge, has never been done before.
%As we shall show here, on the basis of an exact and fully quantum mechanical
%treatment of the light-matter interaction, thermal light can indeed
%induce excited-state coherence in atomic and molecular systems.

We employ an exact (non-perturbative) dynamical
treatment for the interaction of a material system (atomic and molecular) with
a thermal state of the electromagnetic field, where both the material and light subsystems
are quantized. Concretely, we treat the light-matter
system as a bipartite composite system whose subsystems interact via a Jaynes-Cummings-type \cite{jc}
interaction model. We thus express the initial system by
its density matrix, ${\rho(0) = \rho_{F}(0) \otimes \rho_{M}(0)}$
(where $F$ and $M$ stand for `field' and `material', respectively),
and fully propagate it in time according to ${\rho(t)=U(t)\rho(0)U^{\dagger}(t)}$,
where $U(t)$ is the system propagator.
In the case we consider, the material subsystem is initially in its
ground state, while
the field is in a single-mode statistical mixture of Fock states.

As we describe below in detail, we apply the interaction model
to a three-level $V$-type system as a basic model,
and, for the first time to the best of our knowledge, to a two-state Born-Oppenheimer molecular system
for which we propagate the full density matrix, represented in the
electronic $\otimes$ bond-coordinate $\otimes$ photon-Fock
product space.
To conform with the conditions of a natural process in our simulations,
we consider the weak-field
regime and initial average photon number characteristic
of sunlight (about 500nm wavelength at 6000 K, \cite{gerry_knight}).
We find that thermal light induces excited-state coherent dynamics in both
the atomic and molecular
subsystems, as it also does in the regime of strong field and high
average photon number.
We compare this result with that obtained for the interaction
of the material systems with a coherent-state.
Remarkably, we find that both scenarios
exhibit almost identical excited-state coherence features.
On the other hand, in contrast to a coherent-state, thermal light does not induce coherence between the material ground
and excited states.

As a prototype for a molecular system, we consider
a three-level $V$-type model system interacting with a single mode
of the radiation field.
The Jaynes-Cummings-type system Hamiltonian is
\begin{eqnarray}
H &=& \sum_{i} \omega_{i} |i\rangle \langle i| +
\omega(\hat{a}^{\dagger}\hat{a}+\tfrac{1}{2}) \nonumber \\
&~&~~~+ \sum_{i\neq g} \lambda_{i} \left( |i\rangle \langle g| \hat{a} + |g\rangle \langle i| \hat{a}^{\dagger} \right ),
\label{Vsys_Hamiltonian}
\end{eqnarray}
where we have invoked the dipole and rotating-wave approximations,
and set $\hbar =1$. The index $i$ stands for the ground and two
excited states of the $V$ system, $g, e$ and $f$, respectively.
The operators $\hat{a}^{\dagger}$ and $\hat{a}$ are the photon creation and
annihilation operators, respectively, operating on the photon
states $\{|n\rangle\}$. The parameter $\lambda_{i}$ is the
matter-field interaction constant.
Note that this model excludes any intrinsic coupling between the excited states.

The material system is initially in its ground state
${\rho_{M}(0)=|g \rangle \langle g|}$. For the field we
consider two possible initial states: a coherent state and a thermal state.
A coherent state takes the form
$|\alpha \rangle = e^{-\tfrac{1}{2}|\alpha|^{2}}\sum_{n=0}^{\infty}\frac{\alpha^{n}}{\sqrt{n!}}|n\rangle$,
where $ |\alpha|^{2}=\bar{n} $ is the field average photon number.
The corresponding density matrix is $\rho_{F}(0) = |\alpha \rangle \langle \alpha|$.
For the thermal state ${\rho_{F}(0) = \sum_{n}p_{n}|n \rangle \langle n|}$, with
$p_{n} = \tfrac{\bar{n}^{n}}{(1+\bar{n})^{n+1}}$, and the average photon number $\bar{n}$,
given by the Boltzmann distribution \cite{gerry_knight}.
Thus, the initial state of the composite system
is $\rho(0) =\rho_{F}(0) \otimes\rho_{M}(0)$, and
the state of the system at any time in its evolution is given by
\begin{eqnarray}
\rho(t+\Delta t) = e^{-iH \Delta t}\rho(t)e^{iH \Delta t}.
\label{ro_t_Vsys}
\end{eqnarray}
The state of each subsystem is
obtained by tracing $\rho(t)$ over the space of the other subsystem;
that is,
\begin{eqnarray}
\rho_{A}(t) = {\rm Tr}_{B} \left [ \rho(t) \right ].
\label{reduced_ro_t_Vsys}
\end{eqnarray}
In the supplemental material, we give an analytical expression for
$\rho(t)$ for the $V$ system, as well as for the coherence element of the
material subsystem for the fully resonant case where $\omega - \omega_{e,f} = 0$.
Furthermore, we compare the analytical and numerical calculations
in order to justify the numerical simulations presented here.

\begin{figure}[h!]
\vspace {-0.5cm}
\begin{center}
\includegraphics[width=9cm]
{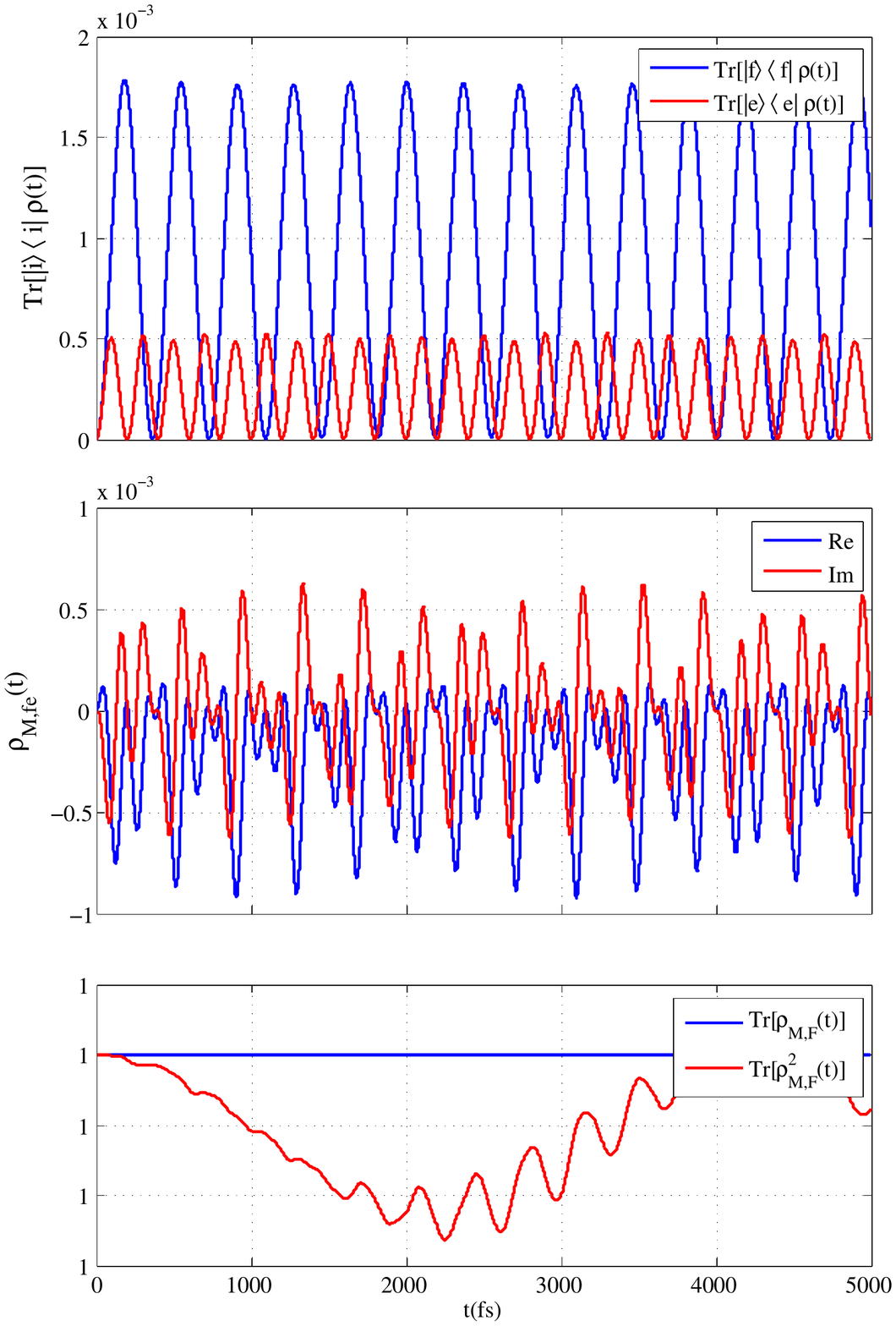}
\end{center}
\begin{center}
\vspace {-1.5cm}
\caption{ \footnotesize{(color online). Populations (top), coherences (middle) and partial traces (bottom) obtained for the interaction of a coherent state
with a three-level $V$-type system.}}
\label{coherent_L12_00001_L13_00001_250_detun_avrg_n_FIX}
\end{center}
\end{figure}

\begin{figure}[h!]
\vspace {-0.5cm}
\begin{center}
\includegraphics[width=9cm]
{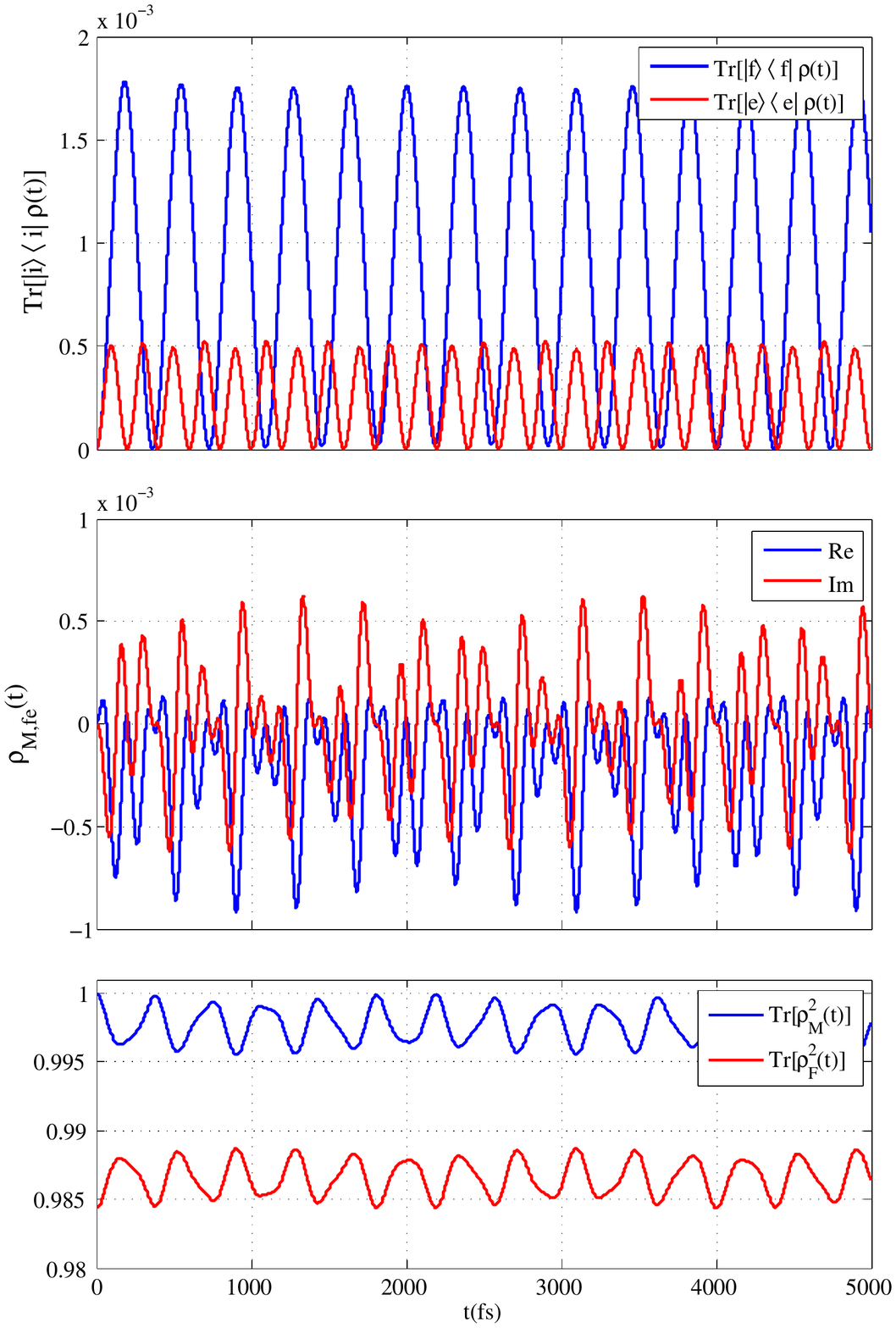}
\end{center}
\begin{center}
\vspace {-1.5cm}
\caption{ \footnotesize{(color online). Populations (top), coherences (middle) and partial traces (bottom) obtained for the interaction of thermal light
with a three-level $V$-type system.}}
\label{thermal_L12_00001_L13_00001_250_detun_avrg_n_FIX}
\end{center}
\end{figure}

We now compare the $V$ system dynamics induced by a coherent
state with that induced by a thermal state. For
both scenarios, the energy separation
between the $|e\rangle$ and $|f\rangle$ states is set to 250 ${\rm cm}^{-1}$,
and the field frequency is tuned between them.
We set $\lambda_{e,f} =  10^{-4}$, and
${\bar{n} = 0.008}$ (obtained for 495.9 nm and temperature 6000 K for the field mode). We construct
the initial states of the light with the first ten
Fock states.
In Fig.~\ref{coherent_L12_00001_L13_00001_250_detun_avrg_n_FIX}, we
show the material excited states populations (top panel), the real and imaginary
parts of the material excited states coherence element, $\rho_{M,fe}(t)$
(middle panel), and the trace of the material (and field) reduced density matrix,
${\rm Tr}\left[\rho_{M,F}(t) \right]$, and of its square, ${\rm Tr}\left[\rho^{2}_{M,F}(t) \right]$
(bottom panel), as obtained for the interaction with the coherent state.
In Fig.~\ref{thermal_L12_00001_L13_00001_250_detun_avrg_n_FIX}, we present the same
dynamical measures for the interaction of the $V$ system with thermal
light; in the lower panel, however, we show only
${\rm Tr}\left[\rho^{2}_{M}(t) \right]$ and ${\rm Tr}\left[\rho^{2}_{F}(t) \right]$
which are generally different for a mixed state.

Figures~\ref{coherent_L12_00001_L13_00001_250_detun_avrg_n_FIX}
and \ref{thermal_L12_00001_L13_00001_250_detun_avrg_n_FIX}
show striking similarity between the material populations and, most remarkably, the excited
state coherence for the coherent and thermal light scenarios.
In this context, it is interesting to refer to the similarity in the interaction of
a coherent and thermal states with a two-level system as noted by Cummings \cite{cummings}.
In contrast,
${\rm Tr}\left [\rho_{M}^{2}(t)\right ]$
and ${\rm Tr}\left [\rho_{F}^{2}(t)\right]$ show
different behavior. In the mixed state scenario (interaction
with the thermal light), the two
oscillate out-of-phase (as if purity is exchanged between the subsystems), while in the pure scenario (interaction with
the coherent state) both are identical, as is well known for a pure
system.
We note that throughout the interaction of the thermal light
with the $V$ system, $\rho_{F}(t)$ remains diagonal as
we show analytically in the supplemental material.

Another major difference between thermal and coherent state of the light
is that while the latter induces coherence between the material ground and excited states,
the former does not. (We derive this result
analytically in the supplemental material).
We explain this phenomenon in the following way. According to our
model, the interaction of a single Fock state, say $|n\rangle$, with
the material ground-state $|g\rangle$ produces the general superposition
$a|g,n\rangle + (b|e\rangle + c|f\rangle)|n-1\rangle$, which in the material subspace
(that is, after tracing over the field subspace) corresponds to a mixed state of the material ground and excited states.
However, the material excited-states are in a coherent superposition with coherences between them.
Thermal light is a statistical mixture of single Fock states and, therefore,
induces only excited-state coherences in the material subsystem.
The absence of coherence between the ground and excited states is related to the vanishing
of the material dipole moment in this scenario as pointed out in \cite{sanchez}.
In contrast, since a coherent state is a {\it coherent} superposition
of Fock states, its interaction with the material ground-state will generate
coherence between the material ground and excited states (since
they share common photon states), and also among the excited states themselves.
Note that a Schr\"{o}dinger cat (coherent) state \cite{gerry_knight} is expected to
behave just like a thermal state, as noted elsewhere \cite{milburn}.

The material excited-state coherence, induced by the thermal light,
is obtained through an exact quantum mechanical treatment without any
intrinsic coupling between the excited states. Such couplings
were introduced in a number of previous studies, and were necessary to allow
the material excited-state coherence. Our exact treatment proves such coherence
is obtained naturally via the interaction with thermal light, similarly to that
obtained in the interaction with a coherent state. In the supplemental material
we show that similar dynamical characteristics are obtained for the interaction of a
the $V$ system with two-mode thermal (and coherent) state.

A realistic molecular system is by far more significant for
understanding matter interaction with thermal light.
We therefore apply a similar fully exact quantum mechanical treatment to a molecular system of
two Born-Oppenheimer potential energy surfaces (PESs).
Specifically, the field is represented in Fock space,
and the molecular system is spanned by the two-dimensional electronic space and
the continuous coordinate space for the nuclei separation.
Thus the state of the system is given by the tensor product of these three spaces.

The Hamiltonian of the entire system is
\begin{eqnarray}
H_{mol} = \sum_{i=g,e} H_{i}(\bm r) |i\rangle \langle i| +
\omega(\hat{a}^{\dagger}\hat{a}+\tfrac{1}{2}) \nonumber \\
+ \lambda \left( |e\rangle \langle g| \hat{a} + |g\rangle \langle e| \hat{a}^{\dagger} \right ),
\label{Molsys_Hamiltonian}
\end{eqnarray}
where $H_{i}(\bm r) = V_{i}(\bm r) +T$ is the
nuclear Hamiltonian of the ground and excited electronic states ($|g\rangle$ and $|e\rangle$),
with the PES $V_{i}(\bm r)$ and the kinetic energy
operator $T = -\frac{1}{2m}\nabla^{2} $. The parameter $m$ is the molecular
reduced mass. We denote the (vibrational) eigenstates of $H_{i}(\bm r)$
by $\psi_{i,\nu}(\bm r)$, where $\nu$ is the vibrational quantum number.
In writing the Hamiltonian of Eq.~\ref{Molsys_Hamiltonian} we invoke the
dipole, Condon, and rotating-wave approximations.

We examine the molecular dynamics induced
by excitation with both thermal and coherent states, as before.
The initial molecular (ground) state
is ${\rho_{M}(0) =  |g,\psi_{g,0} \rangle\langle g,\psi_{g,0}| }$, and
the initial state of the field $\rho_{F}(0)$ is either a coherent or thermal state,
as described above.
The initial state of the system is ${\rho(0) = \rho_{F}(0) \otimes \rho_{M}(0)}$,
and its state at any time in the course of interaction is
\begin{eqnarray}
\rho(t+ \Delta t) = e^{-iH_{mol} \Delta t}\rho(t)e^{iH_{mol} \Delta t }.
\label{ro_t_Molsys}
\end{eqnarray}
It is convenient to employ the ``split operator" method \cite{tannor}
in propagating $\rho$, and split the kinetic energy term of the
Hamiltonian from the rest of the terms.
Thus, for each time-step the propagator is
${e^{-iH_{mol}\Delta t} \approx e^{-i T \Delta t}e^{-i\widetilde{H}_{mol}\Delta t}}$,
where $\widetilde{H}_{mol}$ is the full Hamiltonian of Eq.~\ref{Molsys_Hamiltonian}
without $T$.
In practice, we simulate the interaction of thermal light
with a two-state Born-Oppenheimer model system of two one-dimensional Morse-type
PESs, ${V_{i}({\bm r}) = D_{i}(1-e^{-b_{i}({\bm r}-{\bm r}_{i})})^{2}}+T_{i}$
(the potentials parameters, based on these of the Li$_{2}$ molecule, are given in the supplemental material).
The frequency of the exciting field is tuned between the third
and fourth vibrational eigenstates of the excited state.
The initial state $\rho_{F}(0) = \sum_{n}p_{n}|n\rangle \langle n|$
is constructed with the first seven Fock states. We set $\lambda = 10^{-4}$,
and $\bar{n} = 0.0072$ (obtained for 486.1 nm at 6000 K for the field mode).

\begin{figure}[h!]
\vspace {-0.75cm}
\begin{center}
\includegraphics[height = 16cm, width=10cm]
{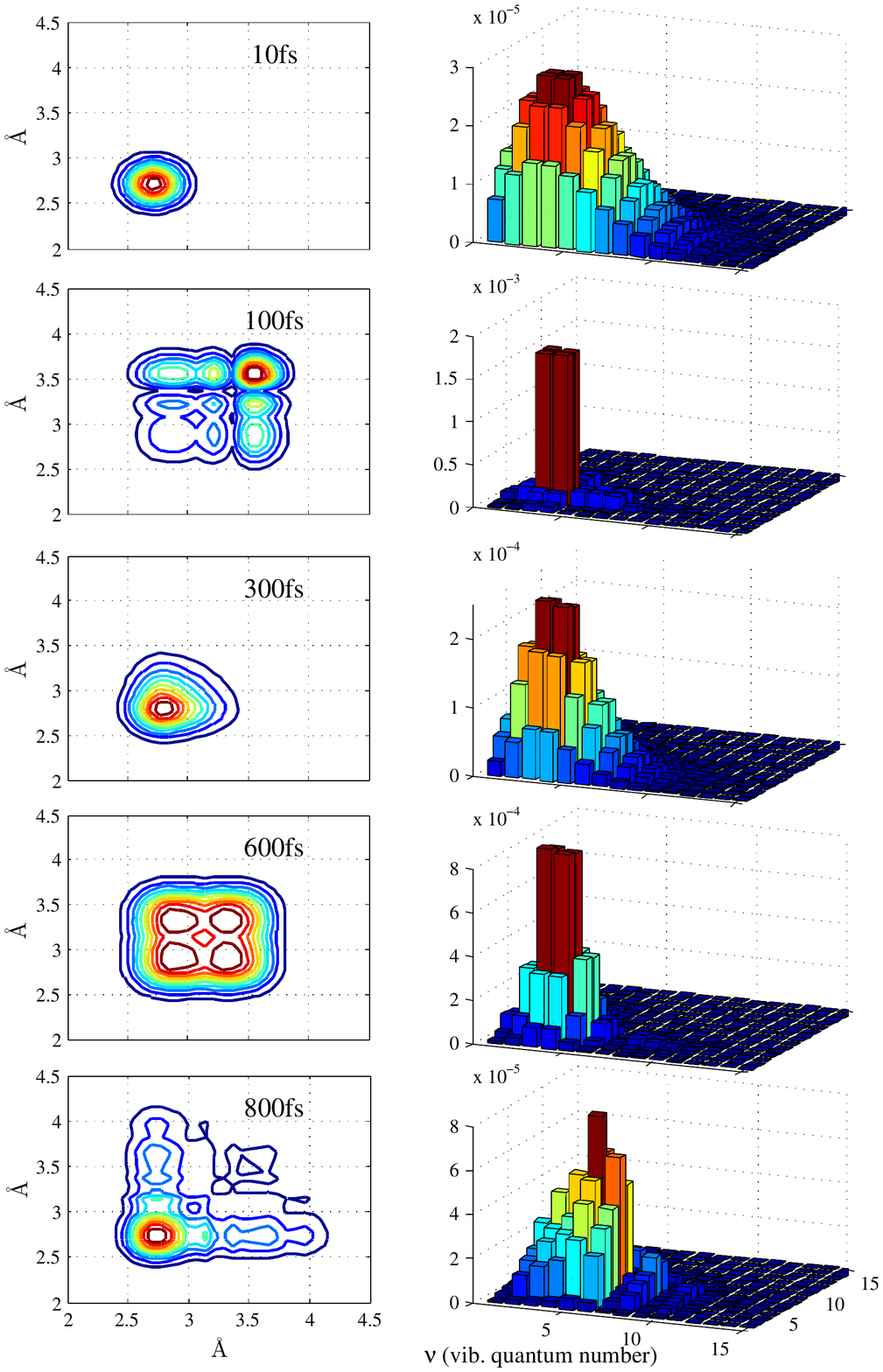}
\end{center}
\begin{center}
\vspace {-1.25cm}
\caption{ \footnotesize{(color online). Snapshots of $\rho_{M, ee}(t)$
for thermal light interacting with two-state Born-Oppenheimer
molecule. The left column shows $\rho_{M, ee}(t)$ in coordinate
space (with the time specified in femtoseconds), and the right column shows its projection onto
the vibrational eigenstates of the molecular excited-state. }}
\label{thermal_ex_st_wavpackt_nFIX}
\end{center}
\end{figure}

In Fig.~\ref{thermal_ex_st_wavpackt_nFIX}, we show snapshots
of the excited-state material reduced density matrix,
$\rho_{M,ee}(t)$. In the left column we show $\rho_{M,ee}(t)$ in
the coordinate representation, while in the right column we show
its projection onto the first 15 excited vibrational
eigenstates.
The clear observation that emerges from these
snapshots is that thermal light induces excited-state vibrational
coherent dynamics.
An almost identical coherence feature is obtained
by the interaction of a coherent state with the molecular system
(not shown here). The spatial representation indicates wavepacket-like
dynamics, unlike that reported recently on the basis of a semiclassical treatment \cite{brumer_2}.
(When the field is resonant with
a vibrational eigenstate of the excited-state,
the state created, both by the interaction with coherent and thermal states,
is the corresponding vibrational eigenstates).
The difference between the interaction with coherent
and thermal state is that the former also induces coherence between
the ground and excited electronic states, while the latter
does not, as discussed above for the $V$ system.
To have a quantitative measure of the wavepacket-like
dynamics, we calculate the excited-state correlation
function ${C(t) = {\rm Tr} \left[ \rho_{M,gg}(0)\rho_{M,ee}(t) \right]}$
and its Fourier-transform $\sigma(\omega)$,
shown in the supplemental material.
The spectral signature is characterized by the harmonics of the energy difference
between the excited-state vibrational eigenstates
$\omega_{ex} \approx 0.0011$ a.u.

The excited-state molecular coherent dynamics, obtained in our
treatment of the interaction with thermal light
contradicts previous studies that either excluded coherence in
the material system, or
allowed the existence of coupling between the excited states that
was necessary for the generation of coherence \cite{shapiro, brumer}. According to our
treatment, the generation of coherence is an intrinsic property
of the interaction between the thermal light and material system.

In conclusion, the question of whether sunlight induces coherent dynamics
in biological systems has been controversial
for many years now, with contradicting answers.
Semiclassical perturbative studies excluded the possibility of
thermal light-induced coherence, while other studies found it possible,
provided there is an intrinsic source that couples the excited
states of the system.

Here we have shown, on the basis of an exact fully quantum mechanical
treatment with a Jaynes-Cummings type of interaction model, and
under conditions that conform with sunlight-induced natural processes,
that thermal light does indeed induce excited state coherent dynamics,
both in a three-level $V$-type system and in a two-state
Born-Oppenheimer molecular systems.
The induced
coherence is similar to that induced by a coherent state.
Remarkably, the thermal light is shown
to induce wavepacket-like dynamics in the molecular system.
Unlike a coherent state, however, the thermal light does not induce coherence
between the material ground and excited states.
Most significantly, and in contrast to previous studies, the generation of the excited-state
coherences in our model requires no intrinsic coupling between the excited states.
In general, thermal light excludes coherence between
the states it couples directly, and permits
coherence between material states which are not coupled directly by the field
or by any other coupling mechanism.
Interestingly, a Schr\"{o}dinger cat (coherent) state
is expected to induce coherences similar to thermal light.
On the basis of recent studies \cite{brumer}, we expect that
the interaction of the system with environment would not
eliminate the induced coherence.
In addition, based on our two-mode calculations, we expect that a continuous-mode
of thermal light, interacting with matter, should induce similar type of coherences.

The results we present in this Letter may suggest a significant role that a
statistical source of modes can have in coherent dynamics, as
was pointed out in a similar context \cite{plenio}.
Consequently, our results may have direct impact
on the understanding the dynamical characteristics of sunlight-induced
biological processes.
\\
\\
$^{\dagger}$Author to whom correspondence should be addressed,
david.avisar@gmail.com

\newpage

\section{Supplementary Material}

\subsection{Analytical dynamical expressions for the interaction
of the $V$ system}

In the interaction picture, the Hamiltonian for the $V$-type
three-level system, assuming $\omega - \omega_{e,f} = 0$ and
a common parameter $\lambda$, is given by
\begin{eqnarray}
H_{I} = \lambda\sum_{i =e,f}  \left( |i\rangle \langle g| \hat{a} + |g\rangle \langle i| \hat{a}^{\dagger} \right ) \nonumber \\
\label{Supl_Vsys_Hamiltonian}
\end{eqnarray}
Employing Taylor expansion, we get the following matrix for the
propagator $e^{-iH_{I}t}$ in the three-dimensional material space
\begin{eqnarray}
e^{-iH_{I}t} =
\begin{pmatrix}
\hat{C}_{f}(t)   & \hat{C}_{fe}(t)  & \hat{S}_{fg}(t)\\
\hat{C}_{ef}(t)  & \hat{C}_{e}(t)   & \hat{S}_{eg}(t)\\
\hat{S}_{gf}(t)  & \hat{S}_{ge}(t)  & \hat{C}_{g}(t)
\end{pmatrix},
\label{Supl_Vsys_propagator}
\end{eqnarray}
with the matrix elements given by
\begin{eqnarray}
\hat{C}_{f}(t) &=& \hat{C}_{e}(t) =  \frac{1}{2}\left[{\rm cos}(\sqrt{2\hat{a}\hat{a}^{\dagger}}\lambda t) +1 \right]
\nonumber \\
\hat{C}_{fe}(t) &=& \hat{C}_{ef}(t) = \frac{1}{2}\left[{\rm cos}(\sqrt{2\hat{a}\hat{a}^{\dagger}}\lambda t) -1 \right]
\nonumber  \\
 \hat{S}_{fg}(t) &=& \hat{S}_{eg}(t) = \frac{-i}{\sqrt{2}}\frac{{\rm sin}(\sqrt{2\hat{a}\hat{a}^{\dagger}}\lambda t)}{\sqrt{\hat{a}\hat{a}^{\dagger}}} \hat{a}
\nonumber  \\
\hat{S}_{gf}(t) &=& \hat{S}_{ge}(t) = \frac{-i}{\sqrt{2}}\frac{{\rm sin}(\sqrt{2\hat{a}^{\dagger}\hat{a}}\lambda t)}{\sqrt{\hat{a}^{\dagger}\hat{a}}} \hat{a}^{\dagger}
\nonumber \\
\hat{C}_{g}(t) &=& {\rm cos}(\sqrt{2\hat{a}^{\dagger}\hat{a}}\lambda t). \nonumber \\
\label{Supl_Vsys_propag_matrx_elem}
\end{eqnarray}
The operations of each of the propagator matrix elements on Fock states
are then
\begin{eqnarray}
\hat{C}_{f}(t)|n\rangle &=& \frac{1}{2}\left[{\rm cos}(\sqrt{2(n+1)}\lambda t) +1 \right] |n\rangle \equiv C_{f,n}(t)|n\rangle
\nonumber \\
\hat{C}_{fe}(t)|n\rangle &=& \frac{1}{2}\left[{\rm cos}(\sqrt{2(n+1)}\lambda t) -1 \right] |n\rangle\equiv C_{fe,n}(t)|n\rangle
\nonumber  \\
 \hat{S}_{fg}(t) |n\rangle &=&  \frac{-i}{\sqrt{2}}{\rm sin}(\sqrt{2n}\lambda t) |n-1\rangle\equiv S_{fg,n}(t)|n-1\rangle
\nonumber  \\
\hat{S}_{gf}(t)|n\rangle &=&  \frac{-i}{\sqrt{2}}{\rm sin}(\sqrt{2(n+1)}\lambda t)|n+1\rangle\equiv S_{gf,n}(t)|n+1\rangle
\nonumber \\
\hat{C}_{g}(t) |n\rangle&=& {\rm cos}(\sqrt{2n}\lambda t)|n\rangle\equiv C_{g,n}(t)|n\rangle. \nonumber \\
\label{Supl_Vsys_matrx_elem_on_fock}
\end{eqnarray}

For the initial state ${\rho(0) = |g\rangle \langle g| \otimes \rho_{F}(0)}$ of the composite system,
the density matrix at any time $t$ of the evolution, in the interaction picture, is
\begin{eqnarray}
\rho_{I}(t) =
\begin{pmatrix}
-\hat{S}_{fg}\rho_{F}(0)\hat{S}_{ge}  & -\hat{S}_{fg}\rho_{F}(0)\hat{S}_{ge}  & \hat{S}_{fg}\rho_{F}(0)\hat{C}_{g}\\
-\hat{S}_{fg}\rho_{F}(0)\hat{S}_{ge}  & -\hat{S}_{fg}\rho_{F}(0)\hat{S}_{ge}   & \hat{S}_{fg}\rho_{F}(0)\hat{C}_{g}\\
-\hat{C}_{g}\rho_{F}(0)\hat{S}_{ge}  & -\hat{C}_{g}\rho_{F}(0)\hat{S}_{ge}  & \hat{C}_{g}\rho_{F}(0)\hat{C}_{g}
\end{pmatrix},~~~
\label{Supl_Vsys_ro_t}
\end{eqnarray}
where we have omitted, for convenience, the explicit time dependence of the matrix elements of
Eq.~\ref{Supl_Vsys_propagator}.

First we derive an expression for the coherence element between the ground
and excited state for the case of a coherent state of the field.
We then show that this element is identically zero for the thermal state
of the field. Let us consider the element $-\hat{C}_{g}\rho_{F}(0)\hat{S}_{ge}$.
For the coherent state of the field we plug $\rho_{F}(0) = |\alpha \rangle \langle \alpha |$
into this element and trace over the field space. As a result, we get
\begin{eqnarray}
\rho_{M,gf}(t)&=&{\rm Tr}_{F} \left[ -\hat{C}_{g}(t)|\alpha \rangle \langle \alpha|
\hat{S}_{ge}(t)\right] \nonumber \\
&=&-\sum_{m} \langle m| \hat{C}_{g}(t)|\alpha \rangle \langle \alpha|
\hat{S}_{ge}(t) |m \rangle \nonumber \\
&=& - \langle \alpha| \hat{S}_{ge}(t) \hat{C}_{g}(t)|\alpha \rangle \nonumber \\
&=&\frac{ie^{-|\alpha|^{2}}}{2} \sum_{n} \frac{\alpha^{n} (\alpha^{\ast})^{n+1}}{\sqrt{n!(n+1)!}}
{\rm sin}\left[\sqrt{2(n+1)}\lambda t \right]\nonumber \\
&~&\times{\rm cos}(\sqrt{2n}\lambda t).
\label{Supl_Vsys_gf_coherence_coherent}
\end{eqnarray}
If we plug into $-\hat{C}_{g}\rho_{F}(0)\hat{S}_{ge}$ the expression for the thermal state of the field,
$\sum_{n}p_{n} |n\rangle \langle n|$, and trace over the field space, the
coherence element $\rho_{M,gf}(t)$ is identically zero (as well as for the $eg$ element):
\begin{eqnarray}
\rho_{M,gf}(t)&=&{\rm Tr}_{F} \left[ -\hat{C}_{g}(t)
 \sum_{n}p_{n} |n\rangle \langle n|
\hat{S}_{ge}(t)\right] \nonumber \\
&=&-\sum_{n,m}p_{n} \langle m| \hat{C}_{g}(t)|n \rangle \langle n|
\hat{S}_{ge}(t) |m \rangle \nonumber \\
&=& -\sum_{n}p_{n} \langle n| \hat{S}_{ge}(t) \hat{C}_{g}(t)|n \rangle =0
\label{Supl_Vsys_gf_coherence_thermal}
\end{eqnarray}

The coherence element $\rho_{M,fe}(t)$ is obtained
by tracing over the field subspace of the matrix element $\rho_{I,fe}(t)$:
\begin{eqnarray}
\rho_{M,fe}(t)&=&{\rm Tr}_{F} \left[ -\hat{S}_{fg}(t)\rho_{F}(0)\hat{S}_{ge}(t)\right]\nonumber \\
&=&-\sum_{m} \langle m| \hat{S}_{fg}(t)\rho_{F}(0)\hat{S}_{ge}(t) |m \rangle.
\label{Supl_Vsys_fe_coherence}
\end{eqnarray}
For a coherent state, we plug $\rho_{F}(0) = |\alpha \rangle \langle \alpha|$
(with ${|\alpha \rangle = e^{-\tfrac{1}{2}|\alpha|^{2}}\sum_{n=0}^{\infty}\frac{\alpha^{n}}{\sqrt{n!}}|n\rangle}$)
into Eq.\ref{Supl_Vsys_fe_coherence} and get the following expression
for the excited-state coherence element
\begin{eqnarray}
\rho_{M,fe}(t) &=& \frac{1}{2} e^{-|\alpha|^{2}} \sum_{n} \frac{|\alpha|^{2n}}{n!} {\rm sin}^{2}(\sqrt{2n}\lambda t)
\nonumber \\
&=&\frac{1}{2} e^{-\bar{n}} \sum_{n} \frac{\bar{n}^{n}}{n!} {\rm sin}^{2}(\sqrt{2n}\lambda t) .
\label{Supl_Vsys_fe_coherence_coherent}
\end{eqnarray}
For the case of thermal field, with $\rho_{F}(0) = \sum_{n}p_{n}|n\rangle \langle n|$, we get
\begin{eqnarray}
\rho_{M,fe}(t)&=&\frac{1}{2}\sum_{n}p_{n} {\rm sin}^{2}(\sqrt{2n}\lambda t) \nonumber \\
&=& \frac{1}{2}\sum_{n} \frac{\bar{n}^{n}}{ (1+\bar{n})^{n+1} }{\rm sin}^{2}(\sqrt{2n}\lambda t).
\label{Supl_Vsys_fe_coherence_thermal}
\end{eqnarray}

In addition, we can also show that during the interaction of the thermal
light with the $V$ system, the reduced density matrix of the light, $\rho_{F}(t)$, remains diagonal.
To show that we trace $\rho_{I}(t)$ of Eq.~\ref{Supl_Vsys_ro_t} over
the material subspace, and use the the equalities of Eq.~\ref{Supl_Vsys_matrx_elem_on_fock}.
The result obtained is
\begin{eqnarray}
\rho_{F}(t)&=& {\rm Tr}_{M} \left[ \rho_{I}(t) \right] = \sum_{i}\langle i|\rho_{I}(t) |i \rangle \nonumber \\
&=& -2\hat{S}_{fg}(t)\rho_{F}(0)\hat{S}_{ge}(t) + \hat{C}_{g}(t)\rho_{F}(0)\hat{C}_{g}(t) \nonumber \\
\nonumber \\
&=& \sum_{n} p_{n} {\rm sin}^{2}(\sqrt{2n}\lambda t)|n-1\rangle \langle n-1| \nonumber \\
&~&~~~~+
\sum_{n} p_{n} {\rm cos}^{2}(\sqrt{2n}\lambda t)|n\rangle \langle n|,
\label{Supl_Vsys_field_ro}
\end{eqnarray}
which represents a diagonal reduced density matrix $\rho_{F}(t)$.

\begin{figure}[h]
\vspace{-3cm}
\begin{center}
\includegraphics[width=8.3cm]
{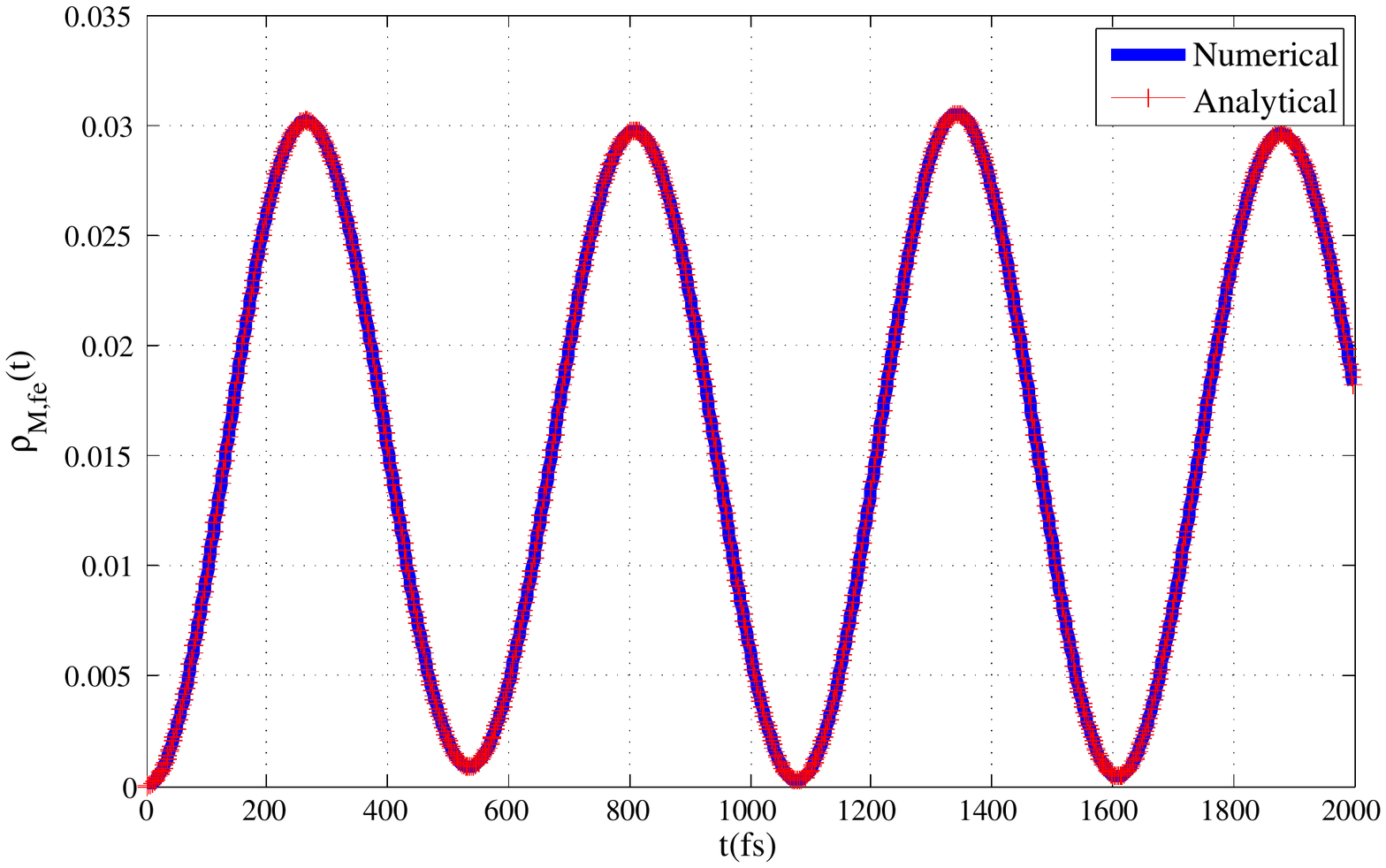}
\end{center}
\begin{center}
\vspace {-3.0cm}
\caption{ \footnotesize{(color online.) Real part of $\rho_{M,fe}(t)$ obtained
numerically, after Eq.~2 of the main text (blue) and analytically, after Eq.~\ref{Supl_Vsys_fe_coherence_coherent},
(red) for the interaction of a coherent state
with a $V$ system. }}
\label{V_pure_analytic_vs_numeric_real_part_coherence_weak_interac}
\end{center}
\end{figure}

\begin{figure}[h]
\vspace{-3cm}
\begin{center}
\includegraphics[width=8.3cm]
{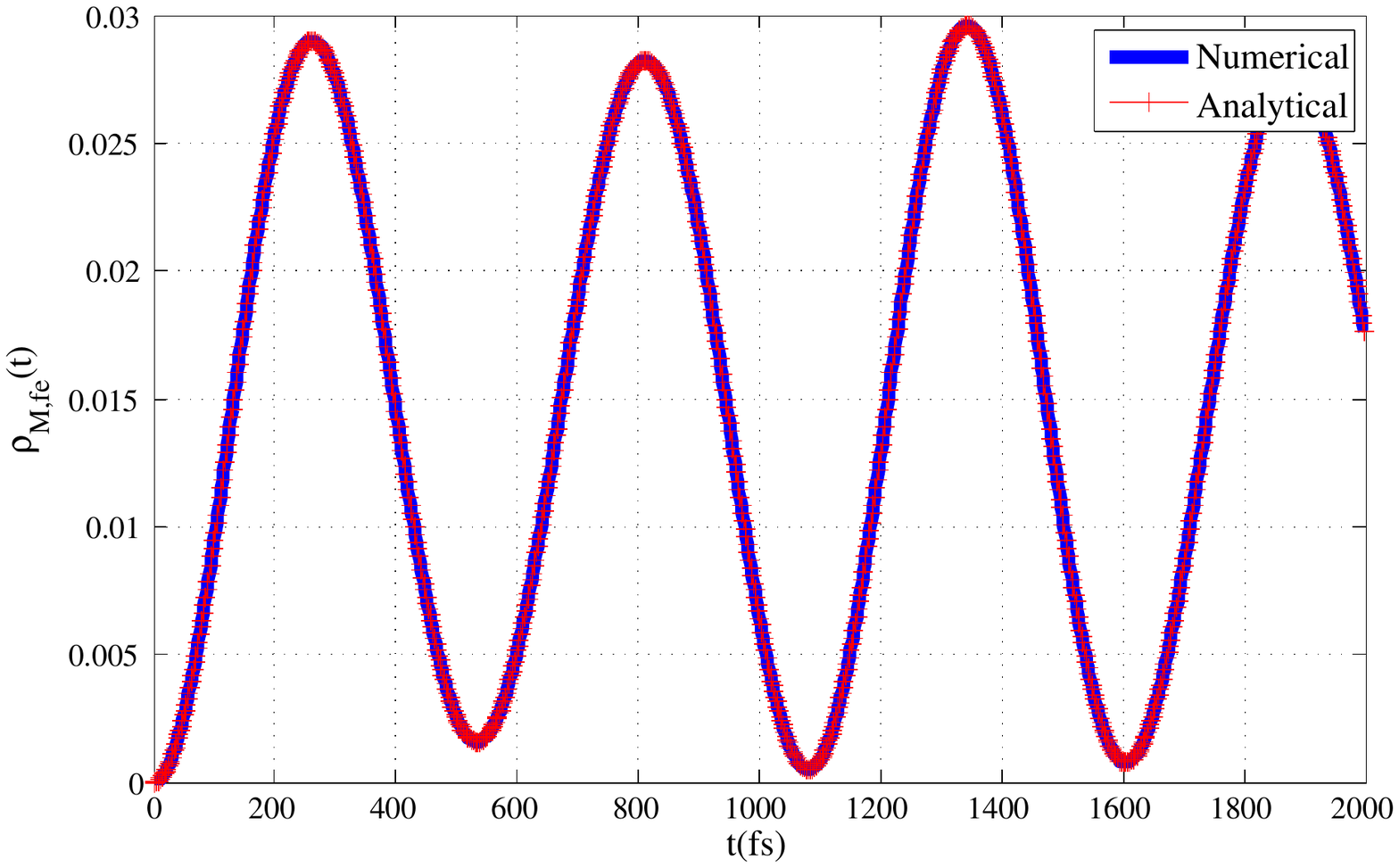}
\end{center}
\begin{center}
\vspace {-3.0cm}
\caption{ \footnotesize{(color online.) Real part of $\rho_{M,fe}(t)$ obtained
numerically, after Eq.~2 of the main text (blue) and analytically, after Eq.~\ref{Supl_Vsys_fe_coherence_thermal},
(red) for the interaction of the thermal light
with a $V$ system. }}
\label{V_thermal_analytic_vs_numeric_real_part_coherence_weak_interac}
\end{center}
\end{figure}

We use the above analytical expressions for the excited states coherence
to compare with our numerical results and verify our numerical
simulations for the density matrix dynamics.
In Fig.~\ref{V_pure_analytic_vs_numeric_real_part_coherence_weak_interac} we show
the real part of the excited state coherence element of the density matrix
$\rho_{M,fe}(t)$,
obtained for the field initially in a coherent state. In blue is the
result obtained from the numerical propagation, whereas in red
is the result obtained using Eq.~\ref{Supl_Vsys_fe_coherence_coherent}.
\vspace{-0cm}
In Fig.(\ref{V_thermal_analytic_vs_numeric_real_part_coherence_weak_interac}), we
show the comparison of the analytical (following Eq.~\ref{Supl_Vsys_fe_coherence_thermal})
and numerical calculation for the
excited state coherence obtained for the thermal field.
Note the remarkable similarity of the results obtained for the coherent
state of the light and the thermal state!
\vspace{-0cm}
Figures (\ref{V_pure_analytic_vs_numeric_real_part_coherence_weak_interac}) and
(\ref{V_thermal_analytic_vs_numeric_real_part_coherence_weak_interac}) confirm
unequivocally our numerical propagation for the density matrix.

\subsection{Interaction of $V$ system with two-mode field.}

Here we show that, similar to the interaction of a singl-mode
states of the light, the interaction with two-mode states of a
coherent state and thermal state, induce excited-state coherence.
\begin{figure}[h!]
\vspace{-1cm}
\begin{center}
\includegraphics[width=7.5cm]
{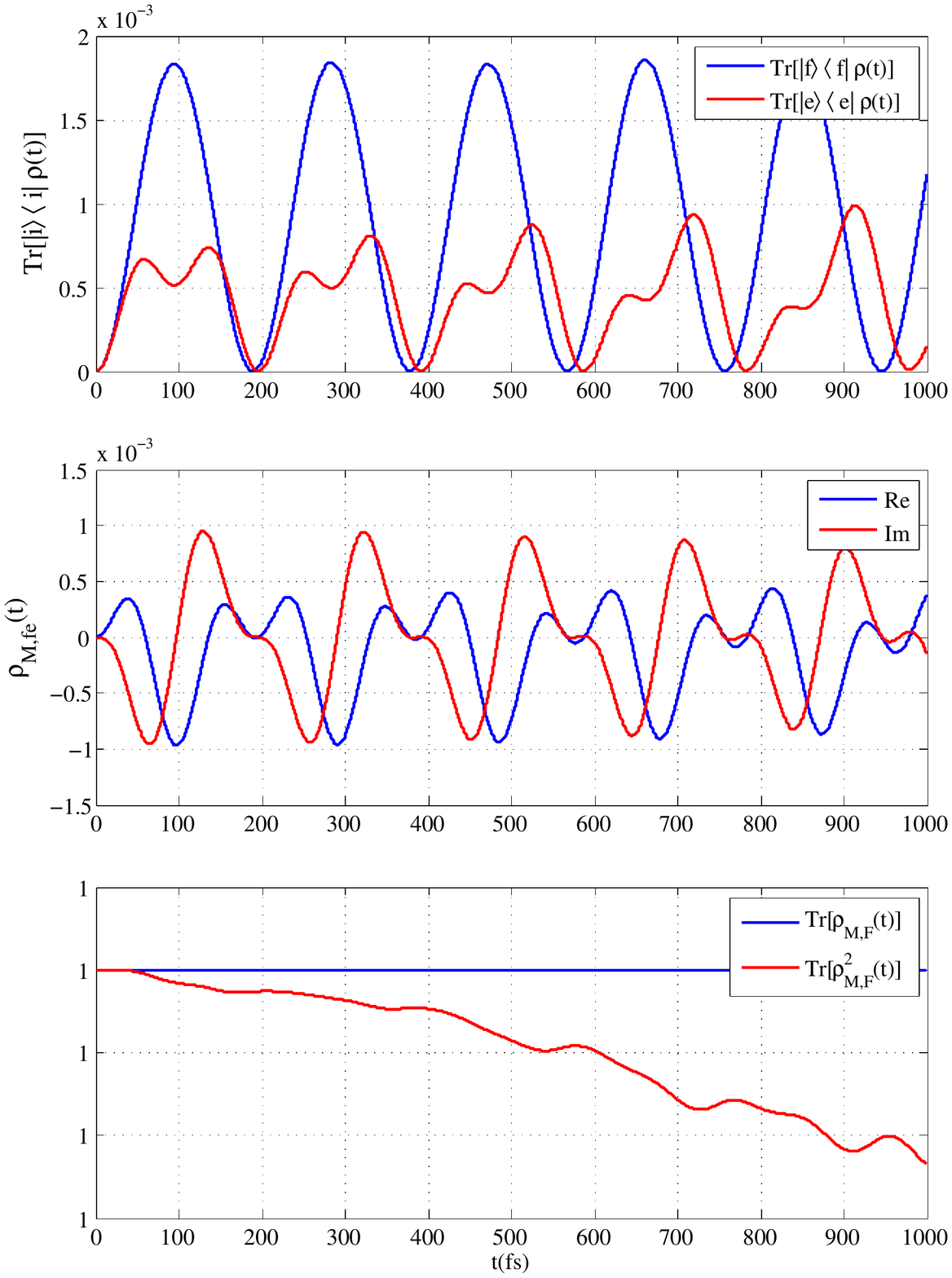}
\end{center}
\begin{center}
\vspace {-1.5cm}
\caption{ \footnotesize{(color online.) Populations (top), coherences (middle) and partial traces (bottom) obtained for the interaction of
a two-mode coherent state with $V$ system.  }}
\label{V_2mode_pure_weak_n_FIX_00001}
\end{center}
\end{figure}

\begin{figure}[h!]
\vspace{-1cm}
\begin{center}
\includegraphics[width=7.5cm]
{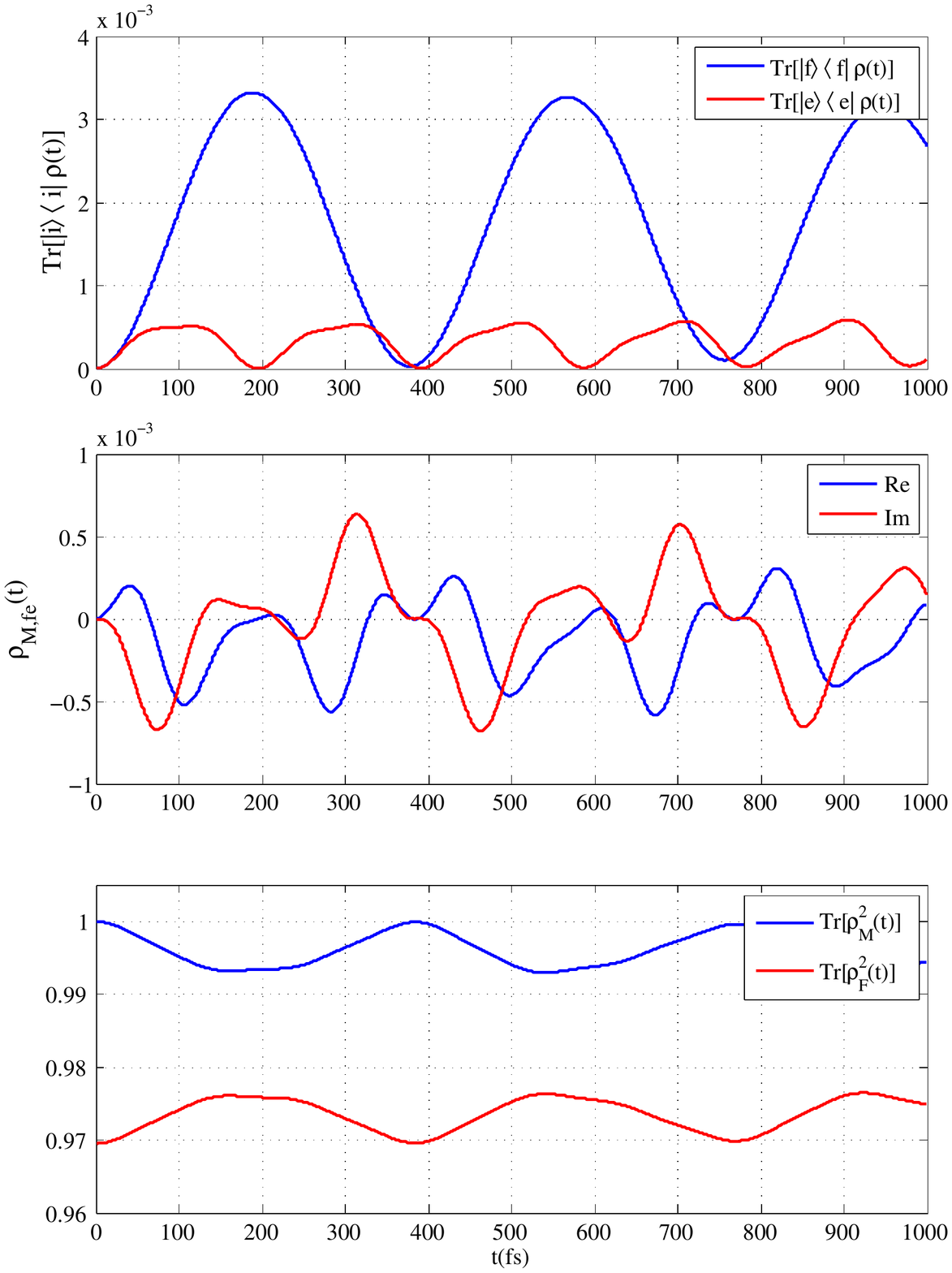}
\end{center}
\begin{center}
\vspace {-1.5cm}
\caption{ \footnotesize{(color online.) Populations (top), coherences (middle) and partial traces (bottom) obtained for the interaction of
a two-mode theraml light with $V$ system.}}
\label{V_2mode_Thermal_weak_n_FIX_00001}
\end{center}
\end{figure}

In Fig.~\ref{V_2mode_pure_weak_n_FIX_00001} we show the dynamical
measures, as discussed in the main text, of the interaction of
a $V$ system with two-mode of a coherent state.
In Fig.~\ref{V_2mode_Thermal_weak_n_FIX_00001} we show the
same dynamical measures obtained for the interaction of
a $V$ system with two-mode thermal state of the light

\subsection{Potentials parameters}

Below are the potentials parameters used for the two-state
Born-Oppenheimer molecular dynamics. The PESs are essentially
the $X$ and $A$ states of the Li$_{2}$ molecule, but only with
the $A$ state shifted in energy \cite{herzberg}.

\begin{table}[h]
\vspace{-0.3cm}
\caption{\label{table_pot} \footnotesize{The parameters,
in atomic units, for PESs used for the molecular
simulations.}}
\begin{ruledtabular}
\begin{tabular}{c @{\hspace{1.5mm}} | @{\hspace{-1mm}} c  c c}
~ &$V_{g}$ & $V_{e}$ \\
\hline
\vspace{-0.3cm}
\\
$D$ & 0.0378492 & 0.0426108
\\
$b$ & 0.4730844 & 0.3175063
\\
${\bm r}_{0}$ & 5.0493478 & 5.8713786
\\
$T$ & 0 & 0.0911267  \\
\end{tabular}
\end{ruledtabular}
\vspace{-0.3cm}
\end{table}

\subsection{Spectrum of the excited-state coherent dynamics}

\begin{figure}[h!]
\vspace{-2cm}
\begin{center}
\includegraphics[width=8.5cm]
{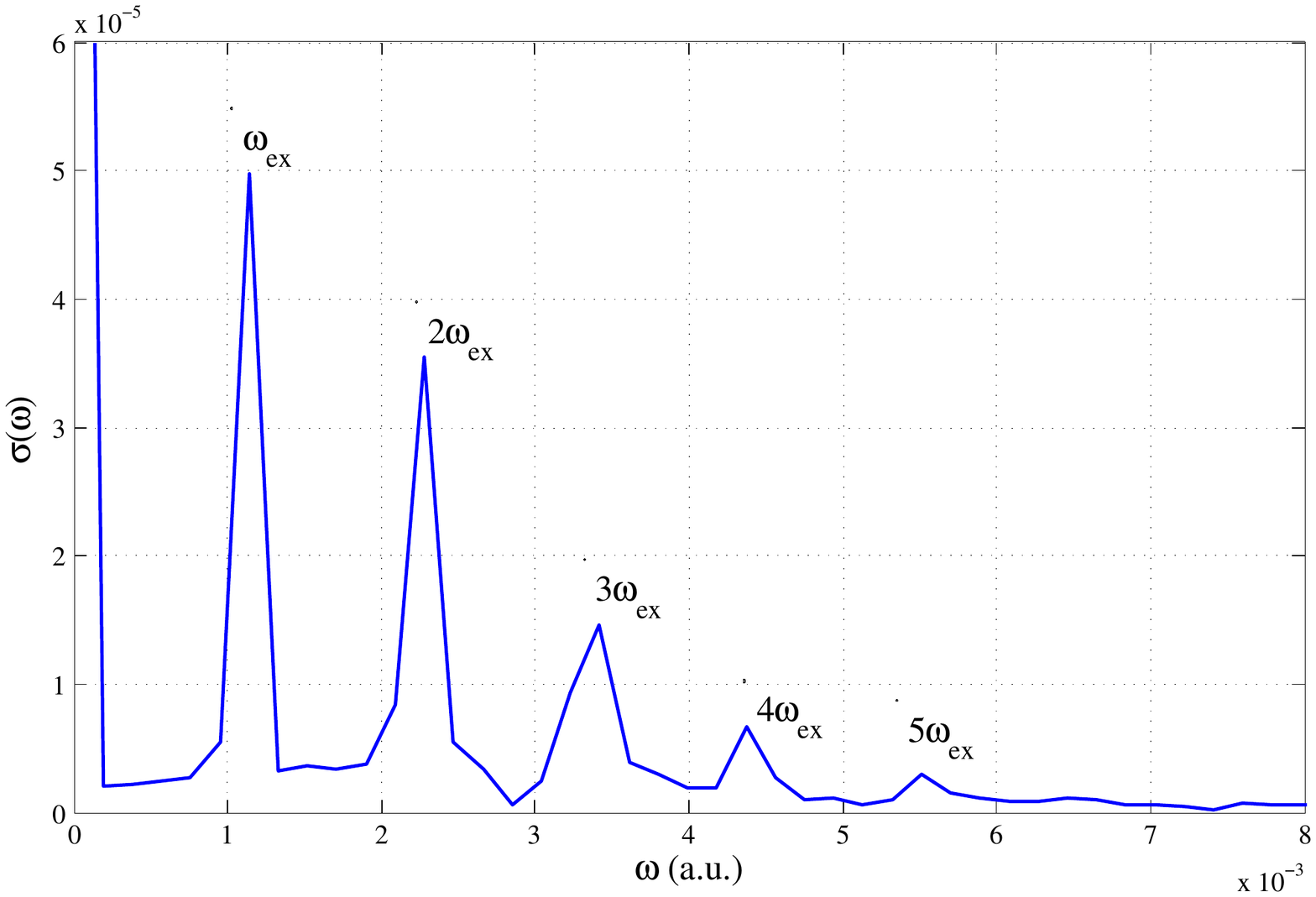}
\end{center}
\begin{center}
\vspace {-3cm}
\caption{ \footnotesize{(color online). The spectrum obtained by Fourier-transform
of the excited-state molecular auto-correlation function ${C(t) = {\rm Tr} \left[ \rho_{M,gg}(0)\rho_{M,ee}(t) \right] }$.}}
\label{mol_thermal_spect}
\end{center}
\end{figure}

In Fig.~\ref{mol_thermal_spect} we show the spectrum obtained
by Fourier-transform of the excited-state correlation function
${C(t) = {\rm Tr} \left[ \rho_{M,gg}(0)\rho_{M,ee}(t) \right]}$.
The spectrum reveals the harmonics of the energy difference
between the excited-state vibrational eigenstates
$\omega_{ex} \approx 0.0011$ a.u.

\end{document}